# YHap: software for probabilistic assignment of Y haplogroups from population re-sequencing data

Fan Zhang[1+], Ruoyan Chen[1+], Dongbing Liu[1], Xiaotian Yao[1], Guoqing Li[1], Yabin Jin[1], Chang Yu[1], Yingrui Li[1*] and Lachlan Coin[1,2,3*]

[1]BGI-shenzhen, China.

[2]Institute of Molecular Bioscience, University of Queensland, Australia.

[3]Department of Genomics of Complex Disease, School of Public Health, Imperial College, London, U.K.

**ABSTRACT**

Y haplogroup analyses are an important component of genealogical reconstruction, population genetic analyses, medical genetics and forensics. These fields are increasingly moving towards use of low-coverage, high throughput sequencing. However, there is as yet no software available for using sequence data to assign Y haplogroup groups probabilistically, such that the posterior probability of assignment fully reflects the information present in the data, and borrows information across all samples sequenced from a population. YHap addresses this problem.

**Availability**: YHap is available from http://www1.imperial.ac.uk/medicine/people/l.coin/

**Contact:** l.coin@imperial.ac.uk ; liyr@genomics.org.cn

## 1 INTRODUCTION

The non-recombining portion of haploid chromosome Y is passed intact from father to son with a mutation rate several times greater than autosomes (Xue et al. 2009). As such, patterns of variation in Y are widely used to uncover historical patterns of human migration; are important in genealogical reconstruction and have application in forensic analyses.

The Y Chromosome Consortium (YCC) published revised Y-chromosome DNA haplogroup tree in 2008, consisting of approximately 600 markers, which can be used to characterize 20 major global haplogroups, labeled A-T, as well as sub-classification into a total of 311 haplogroups at the finest level of resolution. Different major haplogroups have been found at high frequencies in different geographical regions, for example the E clade in Africa, and the O clade in Eastern Asia. Particular fine-level haplogroups are found in multiple locations, such as R1a in Eastern Europe, South Asia and Central Asia, indicating migration of R1a from Eurasian Steppes to the new world. The C3 haplogroup, found at high frequency throughout Asia is commonly interpreted as genealogical remnants of the empire of Genghis Khan 2(Zerjal et al. 2003).

Y haplogroup assignment has traditionally been carried out by targeted genotyping using a combination of short tandem repeat typing, multi-plex PCR and minisequencing (Sanchez et al. 2003, Zerjal et al. 2002), often using a hierarchical strategy in order to first refine the major haplogroup, and subsequently genotype markers within that haplogroup which illuminate finer levels of resolution. Such a procedure requires substantial amount of wet-lab analysis, requires stringent replication and quality control to eliminate errors which can arise due to the limited amount of information collected at each step. More recently, personal genetic companies have included specific Y chromosome markers on custom genotyping arrays (Turner et al. 2008). Nevertheless, the resolution available from genotyping arrays is limited by markers included on the chip.

Very high coverage high throughput sequencing has the potential to capture all single nucleotide and insertion/deletion variation, and as such provide near-perfect assignment of individuals to Y haplogroups. As high coverage sequencing of large population samples remains expensive, low coverage population sequencing, in which each individual is sequenced at less than 2x haploid coverage is an attractive alternative, but this will not capture all individual-level variation. However, given the sharing of haplogroups within an ethnically homogenous population, it should be possible to borrow information across individuals within a population in order to improve haplogroup assignment.

In this application note, we present the YHap tool, which has been primarily designed for assigning haplogroups to low-coverage population re-sequencing data. YHap borrows information across all samples to assign samples to haplogroups probabilistically, thus providing an accurate representation of the inference which can be made from the data collected. YHap is a complete solution and can also be applied to high-coverage sequence data, as well as data from genotyping arrays.

## 2 METHODS

We use the set of haplogroups and mutations defined in (Karafet et al. 2008). We map the forward and reverse primers described in this manuscript to identify the reported strand of the variation in the GRCh37 reference. After strand correction, we identify whether the mutant allele is the equal to the alternative or reference allele, so that we can subsequently work in reference/alternative allele space on the forward strand, consistent with conventional genotype calling schema. Next, we map each mutation to its position on the pre-defined Y phylogenetic tree T. Finally we create a haplogroup matrix $\mathbf{H}$ of size $N_{ref}*M$ where M is the total number of nodes in T (including leaf and internal nodes) and $N_{ref}$ is the number of pre-defined Y markers. Each entry $\mathbf{H}_{ij} = \{H_{ijg}\}$ is a probability distribution vector expressing the probability of a sampled individual from the clade below node $j$ carries allele g (in this case either the reference or alternate allele). At leaf nodes, this probability vector is either

---

[*]To whom correspondence should be addressed.

[+]The authors wish it to be known that, in their opinion, the first two authors should be regarded as joint First Authors.

{0,1} or {1,0}, and at internal nodes, it is the proportion of descendant leaf nodes with reference or alternate alleles, respectively.

To assign a sequenced individual to a specific haplogroup, we obtained genotype likelihoods at each putative variant site (inclusive of all markers in H) from chrY VCF files of 1000 genome project. This results in a matrix S of size L*N where L is the number of sequenced samples, N is the number of putative variants, and $\mathbf{G}_{ij} = \{G_{ijg}\}$ is a vector of genotype likelihoods. We then generated an augmented **H\*** matrix by adding in extra sites in **G** but not **H** with probability vector $H^*_{ij} = \{0.5, 0.5\}$. The pipeline is similar for genotype data, except that the genotype likelihoods are taken to be either 1, if $\mathbf{G}_{i,j} = g$, or 0 otherwise.

We can calculate the assignment of each individual using

$$P(\mathbf{G}_{\cdot j} | \mathbf{H}^*_{\cdot l}) = \prod_{i=1..N} \sum_{g=\{0,1\}} P(\mathbf{G}_{i,j} | g) * P(g | \mathbf{H}^*_{il}) \quad (1)$$

Where $P(g | \mathbf{H}^*_{il}) = H^*_{ijg}$ and $P(\mathbf{G}_{ij} | g) = \mathbf{G}_{ijg}$

We can then calculate the posterior probability of each haplogroup amongst a set of haplogroups, where prior haplogroup probability distribution $P(\mathbf{H}^*_{\cdot l})$ is set to the uniform distribution,

$$P(\mathbf{H}^*_{\cdot l} | \mathbf{G}_{\cdot j}) = \frac{P(\mathbf{G}_{\cdot j} | \mathbf{H}^*_{\cdot l}) P(\mathbf{H}^*_{\cdot l})}{\sum_{k=1..L} P(\mathbf{G}_{\cdot j} | \mathbf{H}^*_{\cdot k}) P(\mathbf{H}^*_{\cdot k})} \quad (2)$$

By restricting the set of haplogroups considered in equation (2), YHap can be customized to either only assign to within the major haplogroups (A through to T), or all possible haplogroups at the finest level of classification.

While this model is sufficient for assigning Y haplogroups individually, it does not capture shared information between sequenced samples adequately, particularly for low coverage sequencing. Given that a population sample will share individuals from the same haplogroup, and while none of these individuals may have enough depth at informative Y haplogroup markers, there is enough information across the pooled reads from all samples from the same haplogroup. However, we do not know a-priori which samples can be pooled as coming from the same haplogroup.

In order to pool information between samples, we treat the allele probability distribution $\{H^*_{ijg}\}$ at markers present in the sequence data but not present as haplogroup markers, as parameters in our model. We update these parameters using expectation maximization, in which we first calculate the posterior probability assigning each sample j to each haplogroup $l$ using equation 2, and then update the $\{H^*_{ilg}\}$ to reflect the average of genotypes assigned to haplogroup $l$ at position $i$, weighted by this posterior probability of assignment. In this way, the model learns which alleles are characteristic of the pre-defined haplogroups, and is thus able to more accurately assign individuals which may not have good coverage at those sites, but which show similarity to other individuals across the Y chromosome. The probabilities $P(\mathbf{H}^*_{\cdot l})$ are also updated at each step to reflect the proportion of haplogroups assigned in the population.

## 3  RESULTS

We applied YHap to low-coverage sequencing data generated in the pilot phase of the 1000 genome consortium which were also part of the Hapmap project, including 19 YRI, 16 JPT, 21 CEU and 9 CHB samples (Abecasis et al. 2010). Major Y haplogroups have been previously assigned to these samples as part of the Hapmap project. The average sequencing depth of these samples is 1.67X as described in 1000 genome Y chromosome analysis report. Compared to haplogroups previously obtained from the Hapmap project (Altshuler et al. 2010), YHap showed perfect assignment accuracy (Table 1). We also used YHap on the Hapmap combined phase 1,2,3 Y genotype data and obtained the same assignments previously reported with this data.

In order to investigate the ability of YHap to assign finer-level haplogroups we compared YHap results obtained at complete resolution (i.e. considering all haplogroup leaf nodes on the pre-defined Y phylogenetic tree) on both Hapmap genotype data and also 1000 genomes low coverage sequence data (Suppl. Table 1). We see that there is complete concordance at the major haplogroup level, and there is increasing uncertainty in assignment as the resolution of assignment increases, particularly using dense genotype data. We also observe that accuracy remains high amongst those assignments which YHap assigns high confidence.

Finally, in order to investigate the relationshis between sequencing depth and assignment accuracy, we randomly downsampled the original bam files from 1000 genome to 0.6X. For simplicity, we chose JPT to run the test. We see that downsampling increases the uncertainty of assignment (Suppl. Table 2), but YHap accuracy remains high amongst those assignments which are made with high posterior probability. This demonstrates that as the underlying amount of information decreases, YHap is still able to extract inference and accurately represent the uncertainty of this inference.

The total complexity for the whole procedure is $O(N^2T)$, conventionally, when using defaut settings, it will take almost 10min to locate 10~20 individuals and approximately 200 Mb memory.

| Pop | ID | Hapmap | 1KG | NGS | Chip | Pop | ID | Hapmap | 1KG | NGS | Chip | Pop | ID | Hapmap | 1KG | NGS | Chip |
|---|---|---|---|---|---|---|---|---|---|---|---|---|---|---|---|---|---|
| CHB | NA18558 | O | N | N1c1c | N1 | CEU | NA06994 | HI | I1 | I1b1 | I1 | YRI | NA18501 | E3a | E1b1a& | E1b1a | E1b1a |
| CHB | NA18561 | O | O2b | O2b | O2 | CEU | NA07357 | R | R1b1b2 | R1b1b2 | R1b1b2 | YRI | NA18504 | E3a7 | E1b1a | E1b1a | E1b1a |
| CHB | NA18562 | O | O | O3a3b1 | O3a | CEU | NA10851 | R | R | R1b1b2 | R1b1b2 | YRI | NA18507 | E3a7 | E1b1a | E1b1a | E1b1a |
| CHB | NA18563 | O | O2b | O2b | O2 | CEU | NA11829 | HI | I1 | I1b1 | I1 | YRI | NA18516 | E3a | E1b1a | E1b1a | E1b1a |
| CHB | NA18572 | O | O | O3a3b1 | O3a | CEU | NA11831 | R | R | R1b1b2 | R1b1b2 | YRI | NA18522 | E3a | E1b1a | E1b1a | E1b1a |
| CHB | NA18603 | O | O | O2a1a | O2 | CEU | NA11881 | HI | I1 | I1b1 | I1 | YRI | NA18853 | E3a | E1b1a& | E1b1a | E1b1a |
| CHB | NA18605 | O | O | O3a3b1 | O | CEU | NA11994 | R | R1b1b2 | R1b1b2 | R1b1b2 | YRI | NA18856 | E1 | E1 | E1/E2 | E1/E2 |
| CHB | NA18608 | O | N | N1c1c | N1 | CEU | NA12003 | HI | I2b | I2b | I2b | YRI | NA18871 | E3a | E1b1a& | E1b1a | E1b1a |
| CHB | NA18609 | O | O | O3a3b1 | O3a | CEU | NA12005 | R | R1b1b2 | R1b1b2 | R1b1b2 | YRI | NA19098 | E3a | E1b1a | E1b1a | E1b1a |
| JPT | NA18940 | D | D2xD2b1 | D2a | D2a | CEU | NA12043 | R | R1 | R1b1b2 | R1b1b2 | YRI | NA19119 | E3a | E1b1a& | E1b1a | E1b1a |
| JPT | NA18943 | O | O2b | O2b | O2 | CEU | NA12144 | R | R1b1b2 | R1b1b2 | R1b1b2 | YRI | NA19138 | E3a | E1b1a& | E1b1a | E1b1a |
| JPT | NA18944 | D | D2b1 | D2a | D2a | CEU | NA12154 | R | R1 | R1b1b2 | R1b1b2 | YRI | NA19141 | E3a | E1b1a& | E1b1a | E1b1a |
| JPT | NA18945 | O | O | O2b | O3a | CEU | NA12155 | R | R1 | R1a1c | R1a1 | YRI | NA19144 | E3a | E1b1a& | E1b1a | E1b1a |
| JPT | NA18948 | D | D2b1 | D2a | D2a | CEU | NA12716 | R | R1 | R1b1b2 | R1b1b2 | YRI | NA19153 | E3a | E1b1a& | E1b1a | E1b1a |
| JPT | NA18952 | D | D | D2a | D2a | CEU | NA12750 | HI | I1 | I1b1 | I1 | YRI | NA19160 | E3a | E1b1a& | E1b1a | E1b1a |
| JPT | NA18953 | O | O2b | O2a | O2 | CEU | NA12760 | R | R | R1b1b2 | R1b1b2 | YRI | NA19171 | E3a7 | E1b1a | E1b1a | E1b1a |
| JPT | NA18959 | O | O | O3a3b1 | O3a | CEU | NA12762 | R | R1 | R1b1b2 | R1b1b2 | YRI | NA19200 | E3a7 | E1b1a | E1b1a | E1b1a |
| JPT | NA18960 | D | D2xD2b1 | D2a | D2a | CEU | NA12812 | R | R1 | R1b1b2 | R1b1b2 | YRI | NA19207 | E3a | E1b1a | E1b1a | E1b1a |
| JPT | NA18961 | D | D | D2a | D2a | CEU | NA12814 | R | R1 | R1b1b2 | R1b1b2 | YRI | NA19210 | E3a7 | E1b1a | E1b1a | E1b1a |
| JPT | NA18965 | O | O2b | O2b | O2 | CEU | NA12872 | R | R1b1b2 | R1b1b2 | R1b1b2 | | | | | | |
| JPT | NA18967 | D | D2b1 | D2a | D2a | CEU | NA12874 | R | R1 | R1b1b2 | R | | | | | | |
| JPT | NA18970 | D | D2xD2b1 | D2a | D2a | | | | | | | | | | | | |
| JPT | NA18971 | C | C1 | C3 | C1a | | | | | | | | | | | | |
| JPT | NA18974 | C | C1 | C3 | C1a | | | | | | | | | | | | |
| JPT | NA19005 | O | O2b | O2b | O2 | | | | | | | | | | | | |

Table 1: Assignment of individuals included in this study to haplogroups. Hapmap indicates results from Hapmap consortium; 1KG results from 1000 genomes consortium; NGS indicates results from YHap applied to 1000 genomes sequence data; Chip indicates results from YHap applied to Hapmap genotype data. The resolution reported for YHap is that for most of which it achieved > 90% posterior probability. *E1b1a was formerly known as E3a.

## ACKNOWLEDGEMENTS

*Funding*: This study was supported by The National Basic Research Program of China, The Chinese 863 program, The Nature Science Foundation of China, The Shenzhen Municiple Government of China.

## REFERENCES

Xue, Y., Wang, Q., Long, Q., Ng, B.L., Swerdlow, H., Burton, J., Skuce, C., Taylor, R., Abdellah, Z., Zhao, Y., et al. (2009). Human Y chromosome base-substitution mutation rate measured by direct sequencing in a deep-rooting pedigree. Curr Biol *19*, 1453-1457

Zerjal, T., Xue, Y., Bertorelle, G., Wells, R.S., Bao, W., Zhu, S., Qamar, R., Ayub, Q., Mohyuddin, A., Fu, S., et al. (2003). The genetic legacy of the Mongols. Am J Hum Genet 72, 717-721

Sanchez JJ, Borsting C, Hallenberg C, Buchard A, Hernandez A, Morling N. Multiplex PCR and minisequencing of SNPs--a model with 35 Y chromosome SNPs. Forensic Sci Int. 2003;137:74-84.

Zerjal, T., Wells, R.S., Yuldasheva, N., Ruzibakiev, R., and Tyler-Smith, C. (2002). A Genetic Landscape Reshaped by Recent Events: Y-Chromosomal Insights into Central Asia. The American Journal of Human Genetics 71, 466-482.

Ann Turner. SNPs on Chips: A New Source of Data for Y-Chromosome Studies. Journal of Genetic Genealogy 4(1):iii-iv,2008

Karafet TM, Mendez FL, Meilerman MB, Underhill PA, Zegura SL, Hammer MF. 2008. New binary polymorphisms reshape and increase resolution of the human Y chromosomal haplogroup tree. *Genome Res* **18**(5): 830-838.

McKenna A, Hanna M, Banks E, Sivachenko A, Cibulskis K, Kernytsky A,

Abecasis, G. R. *et al.* A map of human genome variation from population-scale sequencing. *Nature* **467**, 1061-1073, doi:10.1038/nature09534 nature09534 [pii] (2010).

Altshuler DM, Gibbs RA, Peltonen L, Dermitzakis E, Schaffner SF, Yu F, Bonnen PE, de Bakker PI, Deloukas P, Gabriel SB et al. 2010. Integrating common and rare genetic variation in diverse human populations. *Nature* **467**(7311): 52-58.

Suppl table 1: Concordance of Hapmap array data and 1000 genomes sequence data

| | | | \multicolumn{11}{c|}{Certainty from 1000 genomes sequence data} | |
|---|---|---|---|---|---|---|---|---|---|---|---|---|---|
| | | | 0 | 0.1 | 0.2 | 0.3 | 0.4 | 0.5 | 0.6 | 0.7 | 0.8 | 0.9 | sum |
| Certainty from Hapmap Chip data | Level 1 | 0.9 | | | | | | | | | | 100 (100) | 100 |
| | | sum | | | | | | | | | | 100 | 0 |
| | Level 2 | 0.4 | | | | | 1.5 (0) | | | | | | 1.5 |
| | | 0.5 | | | | | | | | | | | 0 |
| | | 0.6 | | | | | | | | | | 16.9 (100) | 16.9 |
| | | 0.7 | | | | | | | | | | | 0 |
| | | 0.8 | | | | | | | | | | | 0 |
| | | 0.9 | | | | | | | | | | 81.5 (92) | 81.5 |
| | | sum | | | | | 1.5 | | | | | 98.4 | 0 |
| | Level 3 | 0.3 | | | | | | | | | | 16.9 (0) | 16.9 |
| | | 0.4 | | | | | 1.5 (0) | | | | | | 1.5 |
| | | 0.5 | | | | | | | | | | | 0 |
| | | 0.6 | | | | | | | | | | | 0 |
| | | 0.7 | | | | | | | | | | 3.1 (100) | 3.1 |
| | | 0.8 | | | | | | | | | | 1.5 (100) | 1.5 |
| | | 0.9 | | | 3.1 (0) | | | | | | | 73.8 (96) | 76.9 |
| | | sum | | | 3.1 | | 1.5 | | | | | 95.3 | 0 |
| | Level 4 | 0.1 | | | | | | | | | | 6.2 (0) | 6.2 |
| | | 0.2 | | | | | | | | | | | 0 |
| | | 0.3 | | | | | | | | | | 10.8 (0) | 10.8 |
| | | 0.4 | | | | | 12.3 (0) | 1.5 (0) | | 1.5 (0) | | | 15.3 |
| | | 0.5 | | | | | | | | | | 6.2 (100) | 6.2 |
| | | 0.6 | | | | | | | | | | 3.1 (50) | 3.1 |
| | | 0.7 | | | | | | | | | | 3.1 (100) | 3.1 |
| | | 0.8 | | | | | | | | | | 1.5 (100) | 1.5 |
| | | 0.9 | | | 3.1 (0) | | | | | | | 50.8 (97) | 53.9 |
| | | sum | | | 3.1 | | 12.3 | 1.5 | | 1.5 | | 81.7 | 0 |
| | Level 5 | 0.1 | | | | | | | | | | 12.3 (0) | 12.3 |
| | | 0.2 | | | | | 12.3 (0) | | | | | 4.6 (0) | 16.9 |
| | | 0.3 | | | | | | | | | | 13.8 (11) | 13.8 |
| | | 0.4 | | | | | | 1.5 (0) | | 1.5 (0) | | | 3 |
| | | 0.5 | | | | | | | | | | | 0 |
| | | 0.6 | | | | | | | | | | | 0 |
| | | 0.7 | | | | | | | | | | | 0 |
| | | 0.8 | | | | | | | | | | 23.1 (100) | 23.1 |
| | | 0.9 | | | 3.1 (0) | | | | | | | 27.7 (94) | 30.8 |
| | | sum | | | 3.1 | | 12.3 | 1.5 | | 1.5 | | 81.5 | 0 |

Numbers indicate the percentage of samples which had an assignment with a given lower bound on certainty from 1000 genomes sequence data (horizontal) and Hapmap genotype data (vertical), and different levels. Level 1 indicates main haplogroup assignment (A-T), while levels 2-5 indicate increasing levels of precision. Numbers in brackets indicate the concordance rate between predictions made by sequence and genotype data within each bin.

Suppl table 2: Accuracy and certainty of half-coverage

| Certainty | %age | Accuracy (%) |
|---|---|---|
| 0 | 6.2 | 0 |
| 0.1 | 12.5 | 50 |
| 0.2 | 6.2 | 100 |
| 0.3 | 6.2 | 100 |
| 0.4 | 6.2 | 0 |
| 0.5 | 6.2 | 100 |
| 0.6 | 0 | 0 |
| 0.7 | 12.5 | 100 |
| 0.8 | 12.5 | 100 |
| 0.9 | 31.2 | 100 |
| *Average* | | 81.05 |

Percentage of calls made with certainty lower bound on half-coverage at top level of assignment, and concordance with Hapmap assignment.